\documentclass[times,review]{elsarticle}

\usepackage{hyperref}

\usepackage{graphicx}
\usepackage{caption}
\usepackage{subcaption}

\usepackage{commath}
\usepackage{amsmath}

\usepackage{booktabs}
\usepackage[table]{xcolor}
\usepackage{colortbl}
\definecolor{verylightgray}{gray}{0.95}

\journal{Combustion and Flame}

\biboptions{numbers,sort&compress}

\newcommand{\filt}[1] {
  \overline{#1}}

\def\aj{ AIAA Journal }
\def\cf{ Combust. Flame }
\def\jcp{ J.~Comput. Phys. }
\def\jfm{ J.~Fluid Mech. }
\def\pf{ Phys. Fluids}

\graphicspath{{Figs/}}

\hypersetup{draft}

\usepackage{microtype} 
\usepackage{xcolor}
\usepackage[normalem]{ulem}

\usepackage{gensymb}

\begin{document}

\begin{frontmatter}

\title{Training convolutional neural networks to estimate turbulent sub-grid scale reaction rates}

\author[cerfacs]{C.J. Lapeyre\corref{mycorrespondingauthor}}
\cortext[mycorrespondingauthor]{Corresponding author}
\ead{lapeyre@cerfacs.fr}
\author[cerfacs]{A. Misdariis}
\author[cerfacs]{N. Cazard}
\author[ecp]{D. Veynante}
\author[imft]{T. Poinsot}

\address[cerfacs]{CERFACS, 42 avenue Gaspard Coriolis, 31057 Toulouse}
\address[ecp]{EM2C, CNRS, Centrale-Sup\'elec, Universit\'e Paris-Saclay, 3 rue Joliot-Curie, 91192 Gif-sur-Yvette, France}
\address[imft]{IMFT, All\'ee du Professeur Camille Soula, 31400 Toulouse, FRANCE}

\begin{abstract}
  This work presents a new approach for premixed turbulent combustion modeling based on convolutional neural networks (CNN). We first propose a framework to reformulate the problem of subgrid flame surface density estimation as a machine learning task. Data needed to train the CNN is produced by direct numerical simulations (DNS) of a premixed turbulent flame stabilized in a slot-burner configuration. A CNN inspired from a U-Net architecture is designed and trained on the DNS fields to estimate sub-grid scale wrinkling. It is then tested on an unsteady turbulent flame where the mean inlet velocity is increased for a short time and the flame must react to a varying turbulent incoming flow. The CNN is found to efficiently extract the topological nature of the flame and predict subgrid scale wrinkling, outperforming classical algebraic models. This method can be seen as a data-driven extension of dynamic formulations, where topological information was extracted in a hand-designed fashion.
\end{abstract}


\end{frontmatter}

\section{Introduction}

Deep Learning (DL)~\cite{goodfellow2016deep} is a machine learning strategy at the center of a strong hype in many digital industries. This popularity stems in part from the capacity of this approach to sift efficiently through high-dimensional data inherent in \textit{real world} applications. In conjunction with so-called ``Big Data'', or the access to sensing, storage and computing capabilities that yield huge databases to learn from, some challenges \textit{e.g.} in computer vision~\cite{krizhevsky2012imagenet}, natural language processing~\cite{ferrucci2010building} and complex game playing~\cite{silver2017mastering} have seen dramatic advancements in the past decade.

Originally developed as a model of the mammal brain~\cite{mcculloch1943logical}, Artificial Neural Networks (ANN) have since been optimized for numerical performance, enabling the training of deeper architectures, and eventually putting them at the center of the DL effort. These developments have been traditionally lead by experts in computer cognition, limiting their application to select fields. Modern programming frameworks with high levels of abstraction~\cite{tensorflow2015-whitepaper} have however been made available in the past 3 years, in conjunction with powerful hardware such as GPUs to perform fast training. This has opened the possibility for applications in many other fields, such as physics, where the ``causal'' nature of DL~\cite{lin2017does} suggests that complex patterns could also be sought and learned.  

DL clearly belongs to methods devoted to the analysis of {\it data}. In the field of fluid mechanics and of combustion, where {\it models \it i.e.} the Navier-Stokes equations are known, evaluating the possible impacts of DL is difficult. In this area, what is obviously needed is a mixed {\it models / data} approach. Data-driven strategies are by nature approximations, suggesting significant challenges when used on problems for which deterministic equations are available. The low hanging fruits are therefore expected to be sub-problems where models do not rely on exact equations but on simple closure assumptions. In this field, DL may work better than standard models, notably when the flow topology is known to inform the estimation.

Recent studies applied to turbulent flows~\cite{ling2016reynolds,duraisamy2015new, vollant2017subgrid,maulik2017neural} have shown that Sub-Grid Scale (SGS) closure models for Reynolds Averaged Navier-Stokes (RANS) and Large Eddy Simulation (LES) could be addressed using shallow Artificial Neural Networks (ANN). However, advancements offered by DL methods have mostly stemmed from pattern recognition performed by \textit{deep} Convolutional Neural Networks (CNN)~\cite{goodfellow2016deep}, which are still mostly absent from the fluid mechanics literature, as shown in a recent review~\cite{duraisamy2018turbulence}. Nevertheless, some deep residual networks have been built, and it was shown that they could accurately recover state of the art turbulent viscosity models on homogeneous isotropic turbulence~\cite{beck2018neural}.
 
In the combustion community, the determination of the SGS contribution to the filtered reaction rate in reacting flows LES is an example of closure problem that has been daunting for a long time. Indeed, SGS interactions between the flame and turbulent scales largely determines the flame behavior, and modeling them is an important factor to obtain overall flame dynamics. Many turbulent modeling approaches are based on a reconstruction of the subgrid-scale wrinkling of the flame surface and the so-called \textit{flamelet} assumption~\cite{Poinsot:2011}. Under this assumption, the mean turbulent reaction rate can be expressed in terms of \textit{flame surface area}~\cite{Marble:1977a, Candel:1990}. Indeed, the idea that turbulence convects, deforms and spreads surfaces~\cite{Pope:1988} can be applied to a premixed flame front in a turbulent flow. The evaluation of the amount of flame surface area due to unresolved flame wrinkling is the core of all models based on flame surface areas in the last 50 years~\cite{Poinsot:2011}, both for RANS~\cite{Bray:1977a,Peters:1986,Duclos:1993,Bruneaux:1997} and LES~\cite{Boger:1998,Knikker:2004}.  Recent developments of dynamic procedures~\cite{Wang:2011} have shown that extraction of some topological information could increase the accuracy of models. CNNs could be a natural extension of this approach: multi-layer convolutions can be trained to automatically aggregate multi-scale information to predict the desired output. Opening a new path to evaluate subgrid-scale flame wrinkling would be a break-through for turbulent combustion models.

This paper explores this question and proposes \textit{a priori} tests of a \textit{deep} CNN-based model for the SGS contribution to the reaction rate of premixed turbulent flames. It is organized as follows: in Sec.~\ref{sec:theory}, the theoretical aspects of the study are presented. They are inspired from the context of flame surface density models, but are reformulated in the framework of machine learning algorithms. Sec.~\ref{sec:database} describes the DNS performed to produce the data needed to train the neural network. Sec.~\ref{sec:fsd_with_nn} describes the design, implementation and training procedure of a CNN for the flame surface density estimation problem at hand. The data produced in the previous section is used to train a CNN. Once the training process has converged, this network is frozen into a function that is used on new fields to predict flame surface density in Sec.~\ref{sec:results}. In this last section, the accuracy of this trained network is compared to several classical models from the literature, and the specific challenges of evaluating learning approaches is discussed.

\section{Theoretical modeling}
\label{sec:theory}

\subsection{Flame surface density models}

LES relies on a spatial filtering to split the turbulence spectrum and remove the non-resolved scales. For each quantity of interest $Q$ from a well resolved flow field, the low-pass spatial filter $F_{\Delta}$ with width $\Delta$ yields:
\begin{align}
    \filt{Q(\mathbf{x}, t)} &=
    \int_\mathcal{V} F_{\Delta}(\mathbf{x} - \mathbf{x}') Q(\mathbf{x}', t) \dif \mathbf{x}'
    \label{eq:filter_generic}
\end{align}
where $\filt{\cdot}$ denotes the filtering operation. We will limit this study to perfectly premixed combustion where a progress variable $c$ for adiabatic flows is defined as:
\begin{align}
    c &= \frac{T - T_u}{T_b - T_u} \label{eq:progress_var}
\end{align}
with subscripts $u$ and $b$ referring to unburnt and burnt gases, respectively. A balance equation can be written for $c$~\cite{Poinsot:2011}, by defining a density weighted (or Favre) filtering $\widetilde{Q} = \filt{\rho Q} / \filt{\rho}$ for every quantity $Q$.  Filtering the progress variable equation written in a propagative form (G-equation, \cite{Kerstein:1988a}) assuming locally flame elements gives~\cite{Knikker:2004}:
\begin{align}
    \frac{\partial \overline{\rho}\tilde{c}}{\partial t}
    + \mathbf{\nabla} \cdot (\overline{\rho} \tilde{\mathbf{u}} \tilde{c})
    + \mathbf{\nabla} \cdot (\overline{\rho} \widetilde{\mathbf{u} c} - \overline{\rho} \tilde{\mathbf{u}} \tilde{c}) &=
    \rho_u S_L^0 \overline{\Sigma}
    \label{eq:progress_as_surface}
\end{align}
where the right hand side term incorporates filtered diffusion and
reaction terms into a single $c$-isosurface displacement speed assimilated to laminar flame speed $S_L^0$, and where $\rho_u$ is the fresh gases density. $\overline{\Sigma} = \lvert \overline{\mathbf{\nabla} c} \lvert$ is the generalized \textit{flame surface density}~\cite{Boger:1998}, and cannot be obtained in general from resolved flame surfaces.  Indeed, when filtering $c$ to $\overline{c}$, surface wrinkling decreases, resulting in less total $c$-isosurface.  One popular method to model $\overline{\Sigma}$ is to introduce the \textit{wrinkling factor} $\Xi$ that compares the total and resolved generalized flame surfaces. The right-hand side term of Eq.~\ref{eq:progress_as_surface} is then rewritten as:
\begin{align}
    \rho_u S_L^0 \overline{\Sigma} &=
    \rho_u S_L^0 \Xi \lvert \mathbf{\nabla} \overline{c} \lvert
    \label{eq:wrinking_model} \\
    \text{where} ~~~ \Xi &= \frac{\overline{\Sigma}}{\lvert \mathbf{\nabla} \overline{c} \lvert}
    \label{eq:wrinkling_factor}
\end{align}
Fractal approaches such as introduced by Gouldin~\citep{Gouldin:1989} suggest a relationship between $\overline{\Sigma}$ and $\lvert \mathbf{\nabla} \overline{c} \lvert$ of the form:
\begin{align}
    \filt{\Sigma} &=
    \left(\frac{\Delta}{\eta_c}\right)^{D_f - 2}
    \lvert \mathbf{\nabla} \filt{c} \lvert
    \label{eq:gouldin}
\end{align}
where $D_f$ is the fractal dimension of the flame surface, and $\eta_c$ is the inner cutoff scale below which the flame is no longer wrinkled. The $\eta_c$ length scales with the laminar flame thickness $\delta_L^0$~\cite{Poinsot:1991b,Gulder:1995}.

More recent work, based on flame / vortex interactions and multi-fractal analysis~\cite{Charlette:2002} suggests a different form (modified to recover Eq.~\ref{eq:gouldin} at saturation~\cite{Wang:2011}):
\begin{align}
    \filt{\Sigma}
    &= \left( 1 + \min \left[ \frac{\Delta}{\delta_L^0} - 1, 
    \Gamma_{\Delta} \left( \frac{\Delta}{\delta_L^0},
                           \frac{u'_{\Delta}}{S_L^0},
                           Re_{\Delta} \right) \frac{u'_{\Delta}}{S_L^0}
    \right] \right) ^{\beta}
    \lvert \mathbf{\nabla} \filt{c} \lvert
    \label{eq:charlette}
\end{align}
where $\beta$ is a generalized parameter inspired from the fractal dimension. The  $\Gamma_{\Delta}$ function is meant to incorporate the strain induced by the unresolved scales between $\Delta$ and $\eta_c$. Extensions of this model have also been proposed to compute the parameter $\beta$ dynamically~\cite{charlette2002power, Wang:2011}. From a machine learning standpoint, these all correspond to predicting the same output $\overline{\Sigma}$, but using several input variables: $(\overline{c}, \Delta / \delta_L^0, u'_{\Delta} / S_L^0)$.  More variables could be included to further generalize the approach, \textit{e.g.} information about the chemical state, since the machine learning framework does not require a strict physical formulation.

\subsection{Reformulation in the machine learning context}

Flame surface density estimation can be seen as the issue of relating the input field $\overline{c}$ to a matching output field $\overline{\Sigma}$. Supervised learning of this task can be implemented as follows:
\begin{itemize}
    \item in a first phase, a dataset generated using a DNS is used, where both $\overline{c}$ and $\overline{\Sigma}$ are known exactly.  Models are trained on this data in a supervised manner.
    \item in a second step, the best trained model is frozen.  It is executed in an LES context, where $\overline{c}$ is known but not $\overline{\Sigma}$.
\end{itemize}
Both expressions (\ref{eq:gouldin}) and (\ref{eq:charlette}) are fully local: the flame surface depends only on the local characteristics of turbulence ($u'_{\Delta}$), on the grid size ($\Delta$) and on the laminar flame characteristics ($\delta_L^0$ and $S_L^0$). These functions are of the form:
\begin{align}
    \overline{\Sigma} &= f(\overline{c}, \mathbf{u}, ...) \\
    f&: \mathbb{R}^{k} \mapsto \mathbb{R} \nonumber
\end{align}
where $k$ is the number of local variables considered. A generalized DL approach however could use more data by extracting topological information from the flow. In this study, we investigate the capability of spatial convolution to learn to reconstruct the relevant information and produce a function of the form:
\begin{align}
    \overline{\Sigma}(X) &= f(\overline{c}(X)) , ~X \in \mathbb{R}^{n^3} 
    \label{eq:function_shape_nn}\\
    f&: \mathbb{R}^{n^3} \mapsto \mathbb{R}^{n^3} \nonumber
\end{align}
where $n$ is typically $8-32$, and $X \in \mathbb{R}^{n^3}$ is a cube of $n \times n \times n$ adjacent mesh nodes. The nature of this operation differs from classical subgrid scale models which use only local information to infer the subgrid reaction rate: the CNN explores the flow around each point to construct subgrid quantities. Convolutions are promising for this task for several reasons:
\begin{itemize}
    \item convolutions are an efficient strategy to obtain approximations of any order of derivatives of a scalar field~\cite{pratt2007digital};
    \item flames are not local elements but complex structures that spread over several mesh points. Analyzing these structures using algebraic (pointwise) models~\cite{Colin:2000,Charlette:2002} is challenging. The spatial analysis offered by successive convolutions may enable to better understand the global topology of the flame and therefore permit a better estimation of the unresolved structures;  
    \item recent advances in training convolutional neural networks have lead to a high availability of these methods;
    \item convolutions enable to train models on large inputs \textit{via} parameter sharing. This implies that the parameter $n$ in Eq.~\ref{eq:function_shape_nn} can be high, even though the dimensionality of the problem increases with the cube of $n$. This contrasts with other classical machine learning approaches, which would quickly become impractically large on so many inputs.
\end{itemize}

\section{Building the training database}
\label{sec:database}

\subsection{Direct Numerical Simulations of premixed flames}
\label{sec:dns}

In order to obtain $\lvert \mathbf{\nabla} \overline{c} \lvert$ and $\overline{\Sigma}$ fields needed to train the CNN, two DNS of a methane-air slot burner are used.  Their instantaneous snapshots are treated to produce $c$ and $\mathbf{\nabla} c$, and filtered (see Sec.~\ref{sec:dataset}).

The fully compressible explicit code AVBP is used to solve the filtered multi-species 3D Navier-Stokes equations with simplified thermochemistry on unstructured meshes \cite{Schonfeld:1999a,Selle:2004}.  A Taylor{\textendash}Galerkin finite element scheme called TTGC \cite{colin2000development} of third-order in space and time is used.  Inlet and outlet boundary conditions are treated using an NSCBC approach~\cite{Poinsot:1992b} with transverse terms corrections~\cite{Granet:2010}. Other boundaries are treated as periodic. 

\begin{figure}[h!]
    \centering
    \includegraphics[width=0.5\linewidth]{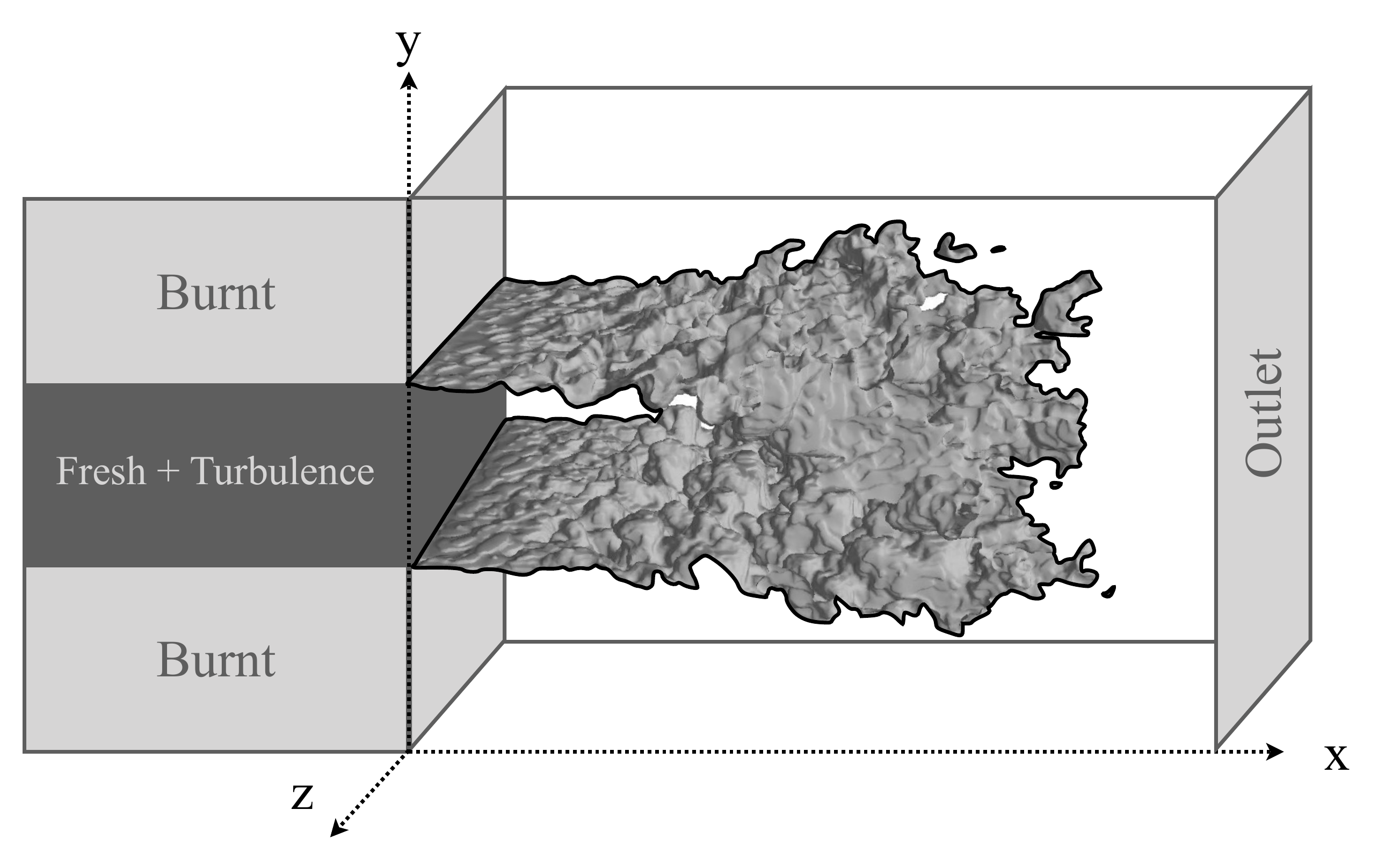}
    \caption{Physical domain used for the DNS.  At the inlet, a double hyperbolic tangent profile is used to inject fresh gases in a sheet $\approx 8$ mm high, surrounded by a slower coflow of burnt gases. Top-bottom (along $y$) and left-right (along $z$) boundaries are periodic. Yellow isosurface is a typical view of $T=1600$ K for DNS2.}
    \label{fig:les_domain}
\end{figure}

Chemical kinetics of the reactions between methane and air at $1$ bar are modeled using a global 2-step scheme fitted to reproduce the flame propagation properties such as the flame speed, the burned gas temperature and the flame thickness~\cite{franzelli2012large}. This simplified chemistry description is sufficient to study the dynamics of premixed turbulent flames. Fresh gases are a stoichiometric mixture with flame speed  $S_L^0 = 40.5$ cm/s and thermal flame thickness $0.34$ mm.
The mesh is a homogeneous cartesian grid with constant element size $\dif x = 0.1$ mm, ensuring that the flame is described on more than three mesh points. Flame speed and thickness were found to be conserved within $5\%$ on a laminar 1D flame.  The domain size is $512$ cells in the $x$ direction and $256$ cells in the $y$ and $z$ ones, for a total of $33.6$ million cells. It is periodic in the $y$ and $z$ directions, and fed by a profile of fresh and burnt gases in the $x=0$ plane (Fig.~\ref{fig:les_domain}).
The inlet is set with a double hyperbolic tangent profile in the $y$ direction, with a central flow of fresh gases enclosed in slower burnt gases coflows. Inlet temperatures are $300$ and $2256$ K in the fresh and burnt gases, respectively. Inlet velocities are $u_{in} = 10$ and $u_{coflow} = 0.1$ m/s.
\begin{itemize}
    \item The central flow is a fresh stoichiometric mixture of methane and air.
    \item The coflow is a slow stream of burnt gases, identical in temperature and mixture to the product of the complete combustion of the central flow.
    \item Turbulence is injected in the fresh gases only. Simulations are performed with either $5\%$ or $10\%$ turbulence injected according to a Passot-Pouquet spectrum~\cite{Passot:1987} with an integral length scale $l_F=2$~mm, yielding $l_F / \delta_L^0 \approx 6$. The fresh gas injection channel has a height $h=8$ mm ($h / \delta_L^0 \approx 25$).
\end{itemize}
Table~\ref{tab:inlet_params} describes the two DNS simulations performed in this study and used to train the CNN.
\begin{table}[h]
  \centering
  \begin{tabular}{@{}rrrrr@{}} \toprule
    & $u_{rms} / \overline{u}$ & Snapshots & $u' / S_L^0$ \\ \midrule
    DNS1 & $5\%$ & 50 & $1.23$\\
    \rowcolor{verylightgray} DNS2 & $10\%$ & 50 & $2.47$ \\
    \bottomrule
  \end{tabular}
  \caption{Parameters for the two DNS simulations performed to produce training data for the CNN.}
  \label{tab:inlet_params}
\end{table}
DNS1 and DNS2 are steady-state simulations, run for $14$~ms each.  The first $4$~ms are transient and discarded, leading to 2 datasets of $10$ ms each, with a full field saved every $0.2$~ms.  This ensures that the fresh gases have traveled approximately $20$ mesh points between each snapshot, yielding significant changes in flame shape and therefore diversity in the training data for the CNN.

\subsection{Dataset}
\label{sec:dataset}

Two meshes are used in this study:
\begin{itemize}
    \item a DNS mesh used to perform the reactive simulations, which contains $512 \times 256 \times 256$ cells.
    \item an ``LES'' mesh, which represents the same domain but 8 times coarser in every direction, \textit{i.e.} $64 \times 32 \times 32$ cells.
\end{itemize}
Fine solutions are produced on the DNS mesh using the Navier-Stokes solver, and then filtered according to Eq.~\ref{eq:filter_generic} and downsampled on the lower resolution LES mesh.
In order to perform this filtering operation, a Gaussian filter is implemented. Its \textit{width} is defined as the multiplying factor on the maximum gradient $\lvert \nabla c \lvert$, \textit{i.e.}:
\begin{align}
    \Delta &= \frac{\max \lvert \nabla c \lvert}{\max \lvert \nabla \overline{c} \lvert} \dif x
\end{align}
computed on a 1D laminar DNS. The resulting function is therefore written in discrete form as:
\begin{align}
    F_{\Delta}(n) &=  \begin{cases}
                         e^{- \frac{1}{2} (\frac{n}{\sigma})^2} & \text{if} ~ n \in [1, N] \\
                         0 & \text{otherwise}
                       \end{cases}
\end{align}
and then normalized by its sum $\sum_{n \in [0, N]} F_{\Delta}(n)$. Here, $\sigma = 26$ and $N = 31$ are optimized to obtain a filter width $\Delta = 8 \dif x \approx 2.3 ~ \delta_l^0 \approx l_F / 2.5$.

Data is often normalized when dealing with machine learning tasks, \textit{e.g} by subtracting the mean and dividing by the standard deviation of the dataset.  In the context of the methodology presented here, these values are not known \textit{a priori} on a new combustion setup, and only the DNS can yield the information for the output data.  The overarching goal of the approach presented here is to apply the technique to cases where a DNS cannot be performed, hence the need for the network to learn features that are not specifically tailored to a single setup.  To achieve this, the input and target fields must be normalized in a fashion that is reproducible \textit{a-priori}.

\begin{figure}[h!]
    \centering
    \includegraphics[width=0.5\linewidth]{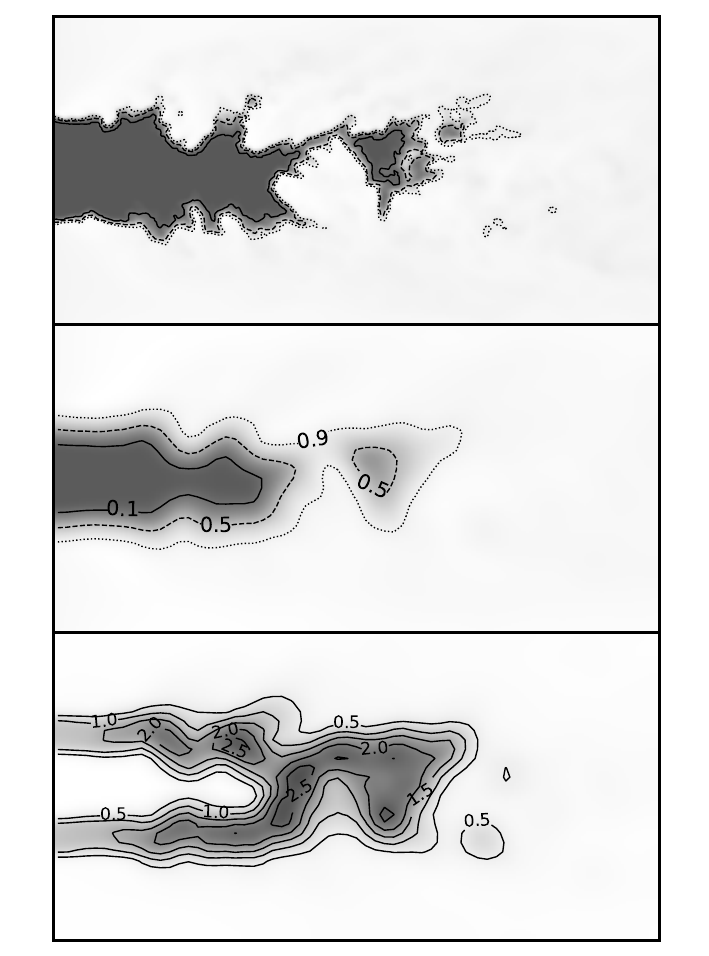}
    \caption{$(x-y)$ slice view of the last field from DNS2 (``snapshot 0`` in Fig.~\ref{fig:dns3_inlet_speed}).  Fully resolved progress variable $c$ (top). From this data, the input of the neural network $\overline{c}$ (middle) and target output to be learned $\overline{\Sigma}^+$ (bottom) are produced.}
    \label{fig:dns_snapshot0}
\end{figure}

\begin{figure}[h!]
    \centering
    \includegraphics[width=0.5\linewidth]{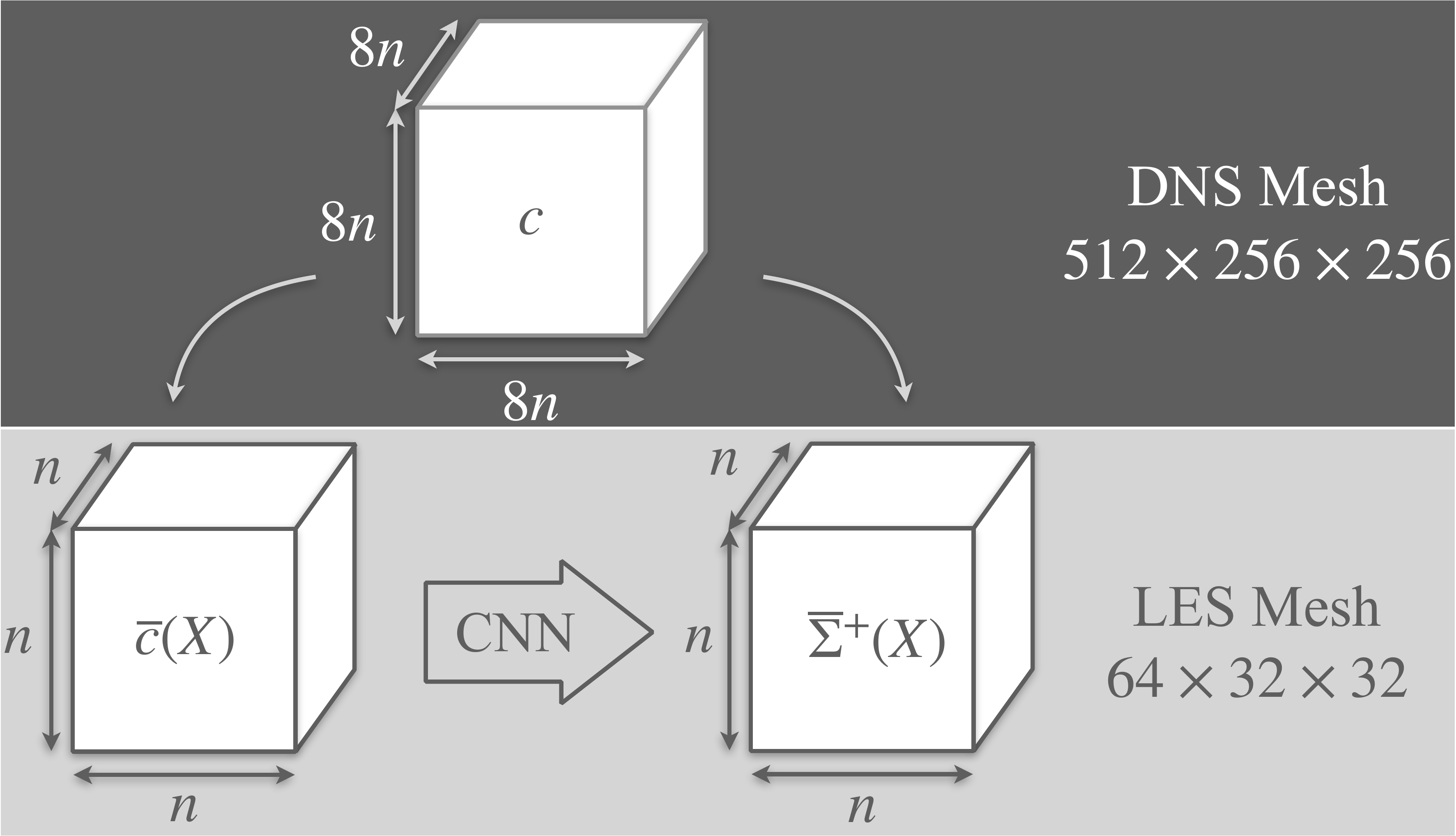}
    \caption{Training strategy to evaluate subgrid scale wrinkling of a premixed flame: the DNS field of $c$ is filtered to produce $\overline{c}$ and $\overline{\Sigma}^+$ on an LES mesh that is 8 times coarser than the DNS. The CNN is then trained on the dataset to approximate the function $\overline{c}(X) \mapsto \overline{\Sigma}^+(X)$ where $X \in \mathbb{R}^{n^3}$.}
    \label{fig:training_strategy}
\end{figure}

To reach this goal, the input field $\overline{c}$ is normalized by construction in Eq.~\ref{eq:progress_var}. Indeed, for premixed combustion this field goes from $0$ in the fresh gases to $1$ in the burnt flow. The output flame surface density value $\overline{\Sigma}$ however spans from $0$ far from the flame (both in fresh and burnt gases) to a maximum value that depends on the amount of subgrid wrinkling of the flame.
The maximum value of $\overline{\Sigma}$ on a laminar 1D flame is used to normalize this field: $\overline{\Sigma}_{lam}^{max}$. The normalized target value writes:
\begin{align}
    \overline{\Sigma}^+ = \frac{\overline{\Sigma}}{\overline{\Sigma}_{lam}^{max}}
    \label{eq:normalized_target}
\end{align}
and does not exceed $1$ in areas where the flame is not wrinkled at the subgrid scale. Values exceeding $1$ suggest unresolved flame surface. Figure~\ref{fig:dns_snapshot0} shows a typical instantaneous snapshot of the configuration in the $(x-y)$ plane: $\overline{\Sigma}^+$ varies between $\approx 1$ near the inlet, where turbulence injection has not yet wrinkled the flame, and a maximum of $\approx 3$ in some local pockets.  This shows how the instantaneous field requires specific FSD estimation locally.
The DNS field is used to produce input and output fields of lower resolution, which in turn are used to train the neural network. The complete training strategy is shown in Fig.~\ref{fig:training_strategy}. The DNS field of $c$ is filtered to produce $\overline{c}$ and $\overline{\Sigma}^+$, then sampled on the $8$ times coarser LES mesh. These two fields are then sampled on $X \in \mathbb{R}^{n^3}$ and fed to the neural network as input/output training.
%

\section{Training the CNN to perform FSD estimation}
\label{sec:fsd_with_nn}
\subsection{Neural network approaches for scalar field generation}
\label{sec:neural_biblio}
The objective of the neural network implementation in this study is as follows:
\begin{itemize}
    \item read on input a 3D field of $\overline{c}(X), ~X \in \mathbb{R}^{n^3} $
    \item train to produce a 3D field of $\overline{\Sigma}^+ (X)$
\end{itemize}
While neural networks were introduced over half a century ago, the recent spike in interest around 2012 was fueled in part by their success on the ImageNet image classification challenge~\cite{krizhevsky2012imagenet}.  Image classification is a task where the input is of high dimension (\textit{e.g.} a 2D, color image), and the output is of lower dimension, typically the number of possible classes ($1000$ in the case of ImageNet).  The network is therefore expected to perform dimensionality reduction, and there is no sense of locality between the input and output.

In this study, a different task is needed: the input field $\overline{c}$ must be mapped at every mesh point with a matching FSD value, yielding a total $\overline{\Sigma}^+$ field of the same dimension as the input.  While traditional machine learning approaches might treat this problem point by point, the neural network can perform an analysis of the full field in a single inference.  In order for this to be efficient, the output must not yield a single node value for $\overline{\Sigma}^+$, but a full field.

This task is closer to a machine learning task called image segmentation: an input image is read and analyzed, and an output of same dimension classifies each pixel.  This task has also gained traction in the community over the same period, notably with the PASCAL-VOC challenge~\cite{everingham2010pascal}.  In this problem, the classification task must be performed without losing the locality of the information.  To this end, fully convolutional neural networks~\cite{long2015fully} were introduced, showing significant progress in segmentation tasks.  Reconstructing the output however took a complex assembly of multiple scales, and later a new architecture which automates this multi-scale analysis called U-Net~\cite{ronneberger2015u} was introduced.  This structure has  been expanded upon, yielding deeper networks that are more accurate for complex segmentation tasks~\cite{chen2018deeplab,taghanaki2018select}.

\subsection{Neural network architecture}
Review of the literature from Sec.~\ref{sec:neural_biblio} suggest that a U-net architecture~\cite{ronneberger2015u} offers a basis for the task at hand in this study.  U-nets however are optimized for 2D image segmentation. This implies that:
\begin{itemize}
    \item the input is expected to be a 3 channel (RGB color) 2D matrix representing an image.  To adapt to the current case, a single channel will be used on input, but as a full 3D scalar field: $\overline{c}(X), ~X \in \mathbb{R}^{n^3}$.
    \item the output of a U-net is meant for a classification task.  Output activation functions are therefore used to represent a categorical distribution, and in the present case must be swapped with rectified linear unit activation to better fit the regression task at hand~\cite{goodfellow2016deep}.
\end{itemize}

\begin{figure}[h!]
    \centering
    \includegraphics[width=0.5\linewidth]{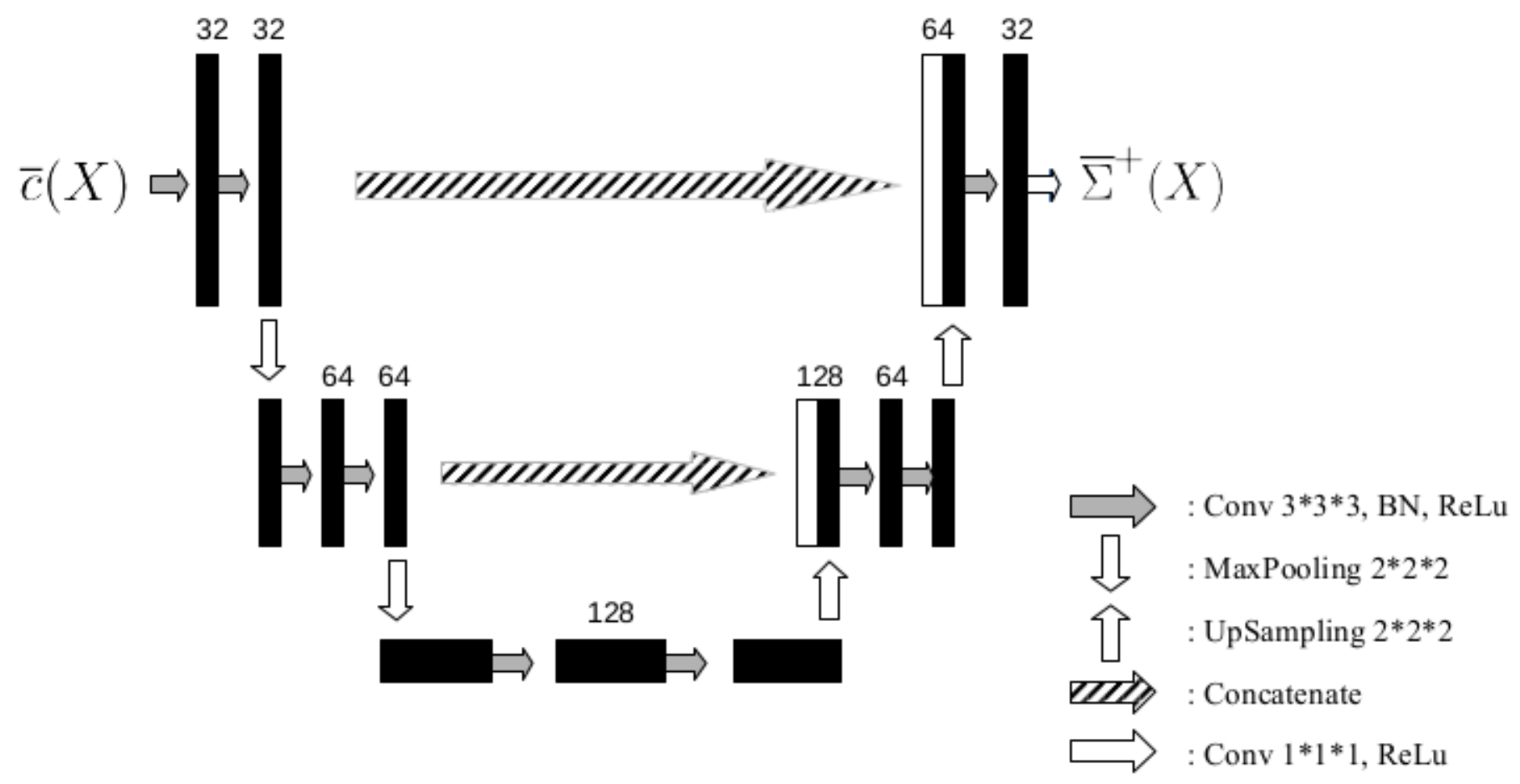}
    \caption{Structure of the U-net - inspired network used in this study. A total of 13 layers are used. Integers above convolutional layers show the number of filters for each convolution.}
    \label{fig:unet_structure}
\end{figure}

The U-net-inspired architecture chosen for this study, adapted for the regression task at hand is shown in Fig.~\ref{fig:unet_structure}.
It consists of a fully convolutional neural network with a downsampling and an upsampling path.  At each downsampling step, two padded 3D convolutions with $3 \times 3 \times 3$ kernel are applied, each followed by a batch normalization (BN) and a rectified linear unit (ReLu) activation.  In addition, a $2 \times 2 \times 2$ maxpooling operation is applied and the number of feature channels is doubled.  The upsampling path is a mirrored version of the downsampling path, with a similar structure: it includes 3D transposed convolutions instead of 3D convolutions, and a $2 \times 2 \times 2$ upsampling operation to recover the initial dimensions.  Additionally, according to the U-net structure, skip-connections link layers with equal resolution of each path.  In order to perform a regression task the final layer, a 3D transposed convolution with $1 \times 1 \times 1$ kernel was used, with a ReLu activation to prevent the network from predicting negative outputs. In total, the network consists of $1,414,145$ trainable parameters, corresponding to all the weights that need to be adjusted in the network. In the following, the network described here is simply referred to as the CNN.

\subsection{Training the CNN}
\label{sec:training_the_CNN}
The data from the two DNS described in Sec.~\ref{sec:dataset} (Tab.~\ref{tab:inlet_params}) is used to train the CNN. In machine learning, the data is classically split in three categories:
\begin{itemize}
    \item the \textit{training} set, used to optimize the weights of the network;
    \item the \textit{validation} set, used to evaluate the error during training on a set that has not been observed. This enables to detect the point where the network starts overfitting to the training set, and additional training starts to increase the error on the validation set;
    \item the \textit{testing} set, kept completely unseen during training, and only used \textit{a posteriori} once the training is converged to assess the performance of the full approach.
\end{itemize}
Training and validation datasets are often taken from the same distribution, and are simply different samples. Ideally, the testing dataset should be taken from a slightly different distribution, in order to show that the underlying features of the data have been learned, and that they can be generalized to new cases. In this study, two DNS with similar setups (DNS1 and DNS2) that lead to similar flames with some variability introduced by different turbulent intensities are used to produce the training and validation sets, by splitting their data (Tab.~\ref{tab:dataset_split}). In order to obtain a testing set from a different distribution, a dedicated simulation DNS3 is performed, as described in Sec.~\ref{sec:results}.
\begin{table}[h]
  \centering
  \begin{tabular}{@{}rccc@{}} \toprule
    & Training & Validation & Testing \\ \midrule
    DNS1 & $1-40$ & $41-50$ & $\varnothing$ \\
    \rowcolor{verylightgray} DNS2 & $1-40$ & $41-50$ & $\varnothing$ \\
    DNS3 & $\varnothing$ & $\varnothing$ & $1-15$ \\
    \bottomrule
  \end{tabular}
  \caption{Data split for the network training and testing in this study. All columns are expressed in terms of DNS snapshot numbers (1 every $0.2$ ms), in sequence.}
  \label{tab:dataset_split}
\end{table}

Additionally, data augmentation during training was found to increase the quality of the results. Each training sample is a random $16 \times 16 \times 16$ crop from the 3D fields, and random $90$\degree rotations and mirror operations are applied since the model should have no preferential orientation and the network must learn an isotropic function.  A training step is performed on a mini-batch of $40$ such cubes in order to average the gradient used for optimization and smooth the learning process. The ADAM~\cite{kingma2014adam} optimizer is used on a mean-squared-error loss function over all output pixels of the prediction compared to the target. $100$ of these mini-batches are observed before performing a test on the validation set to evaluate current train and validation error rates. Each of these $100$ mini-batch runs is called an \textit{epoch}.  The learning rate, used to weight the update value given by the gradient descent procedure, is initially set to $0.01$, and decreased by $20\%$ every $10$ epochs. The network converges in $\approx 150$ epochs, for a total training time of $20$ minutes on an Nvidia Tesla V100 GPU. On this dedicated processor, the dataset is indeed much smaller than typical DL challenge datasets, yielding comparatively short training times.

\section{Using the CNN to evaluate subgrid scale wrinkling}
\label{sec:results}

\subsection{DNS3: a simulation tailored for testing}

\begin{figure*}[h!]
    \centering
    \includegraphics[width=\linewidth]{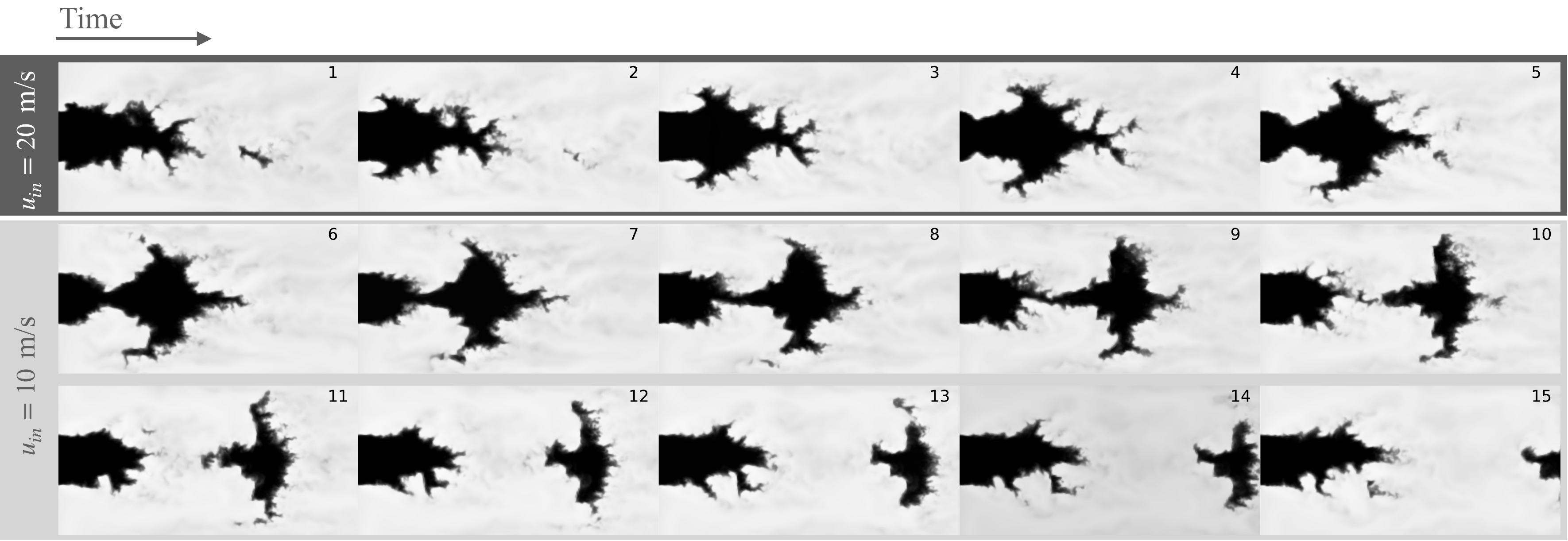}
    \caption{View of $c$ in the $(x-y)$ plane at $z=0$ for all snapshots ($1-15$) of DNS3. Black ($c=0$) to white ($c=1$) shows transition from unburnt to burnt gases, respectively. For this DNS, inlet velocity of the fresh gases is doubled for $1$ ms (5 snapshots), then set back to its original value for $2$ ms (10 snapshots), when the detached pocket of burnt gases reaches the exit.}
    \label{fig:dns3_snapshots}
\end{figure*}

Once the training data has been generated (Sec.~\ref{sec:database}) and the CNN has been fully trained on it (Sec.~\ref{sec:fsd_with_nn}), the network is frozen, and can be used to produce predictions of $\overline{\Sigma}^+$ based on new fields of $\overline{c}$ unseen during training. To verify the capacity of the CNN to generalize its learning, a new, more difficult case (DNS3) was used. DNS3 is a short-term transient started from the last field of DNS2, where inlet velocity is doubled, going from $10$ to $20$ m/s for $1$ ms (5 snapshots), and then set back to its original value for $2$ more ms (Fig.~\ref{fig:dns3_snapshots}).
\begin{figure}[h!]
    \centering
    \includegraphics[width=0.5\linewidth]{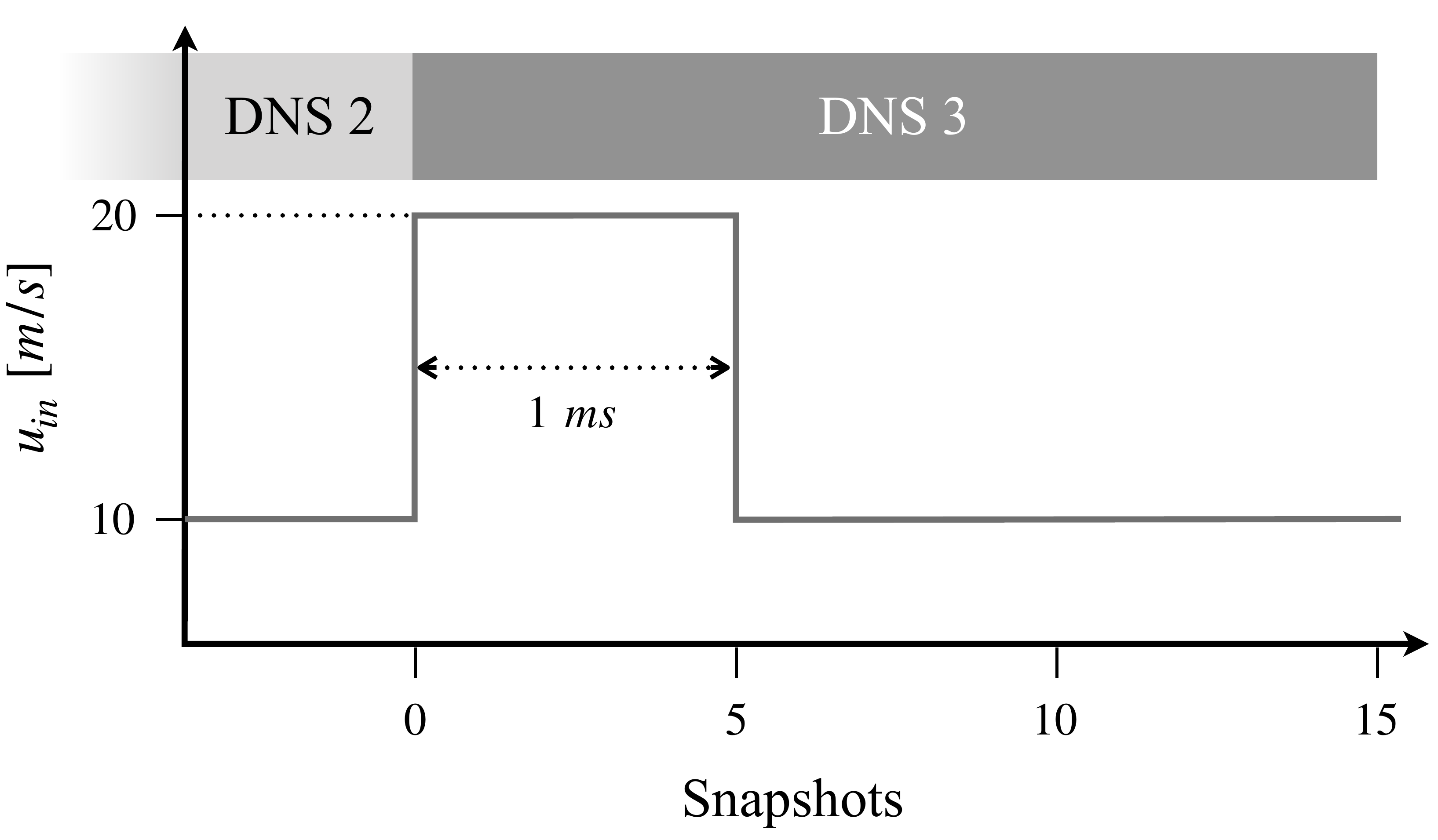}
    \caption{Inlet velocity versus time (1 snapshot every $0.2$ ms) for DNS3, continued from DNS2.}
    \label{fig:dns3_inlet_speed}
\end{figure}
The RMS value of injected turbulence remains constant at $u'=1$ m/s. This sudden change leads to a very different, unsteady flow (Fig.~\ref{fig:dns3_snapshots}) where a ``mushroom''-type structure is generated~\cite{Poinsot:1987c} and where turbulence varies very strongly and rapidly. It is a typical situation encountered in chambers submitted to combustion instabilities, and is now used to evaluate the CNN.

This \textit{a priori} estimation of $\overline{\Sigma}^+(X)$ on new fields of $\overline{c}(X)$ with a trained and frozen network is referred to as \textit{inference} in machine learning, and it is again performed here on the GPU.  Due to the fully convolutional nature of the chosen network, $X$ need not be in $\mathbb{R}^{n^3}$: the network can be directly executed on a 3D flow field of any size. Inference is therefore performed on each full-field snapshot in a single pass. This has the strong advantage that there is no overlapping region between inference areas, in which the predictions can be of poorer quality~\cite{beck2018neural}. Inference time is $12$ ms for each $64 \times 32 \times 32$ LES field observed. 

\begin{figure}[h!]
    \centering
    \includegraphics[width=0.5\linewidth]{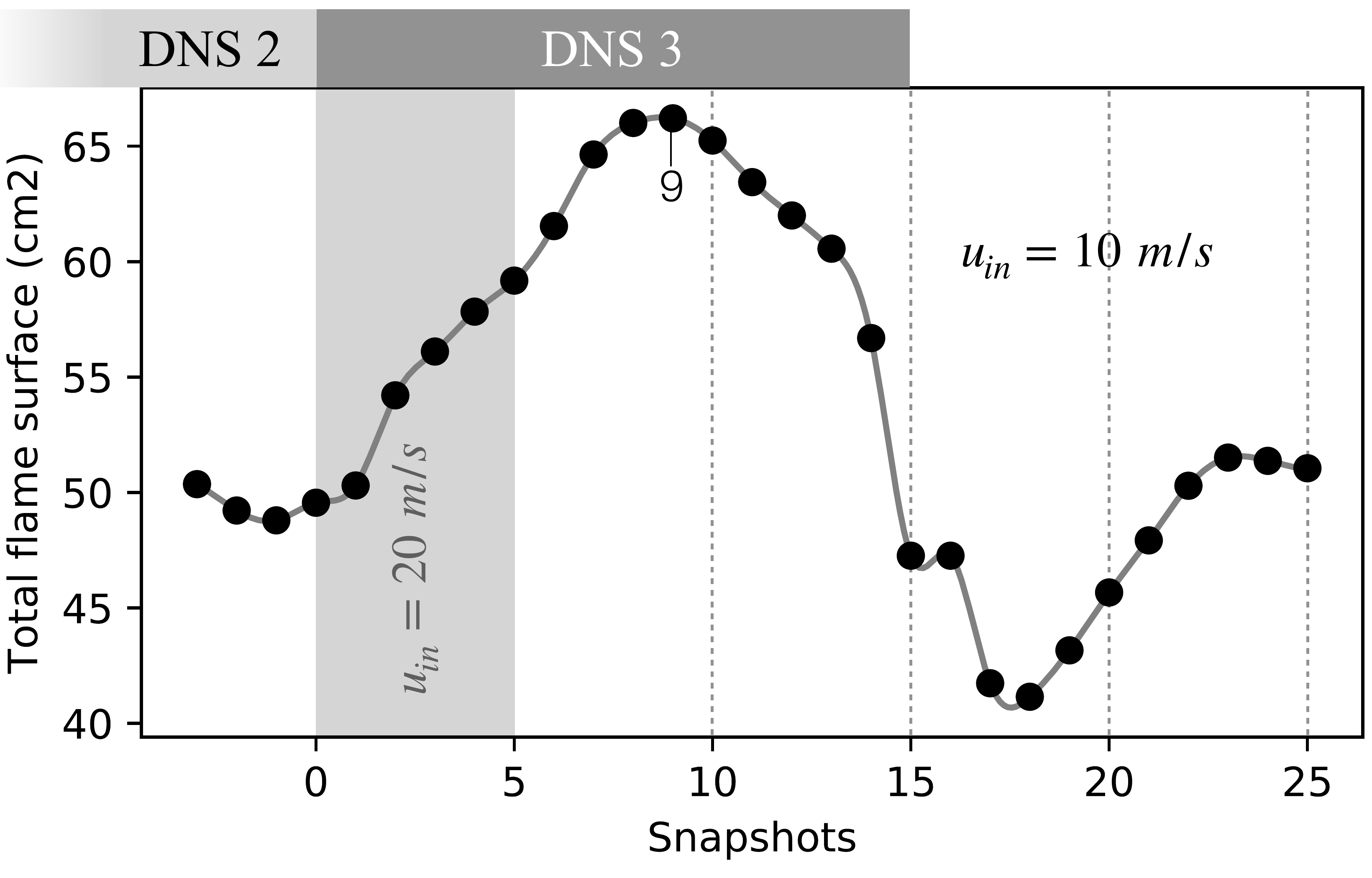}
    \caption{Total flame surface in the domain versus time during DNS3. Test set spans snapshots 1 through 15. A view of the field from snapshot 9 is shown in Fig.~\ref{fig:dns3_snapshot9}.}
    \label{fig:DNS3_flame_surface}
\end{figure}

\begin{figure}[h!]
    \centering
    \includegraphics[width=0.5\linewidth]{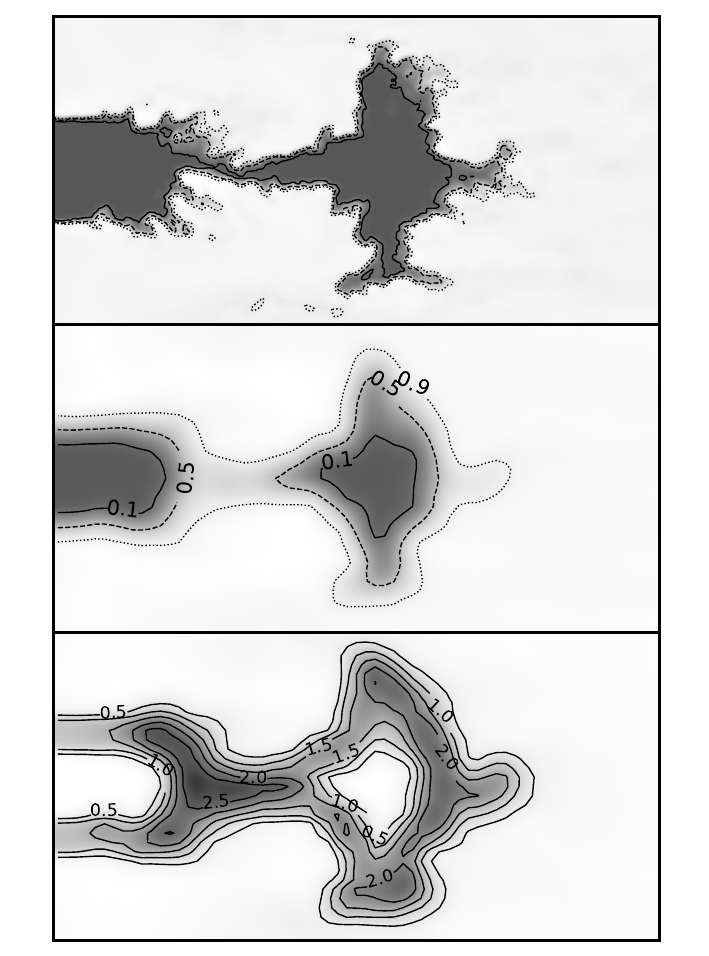}
    \caption{$(x-y)$ slice view of snapshot 9 from DNS3.  Fully resolved progress variable $c$ (top). From this data, the input of the neural network $\overline{c}$ (middle) and target output to be learned $\overline{\Sigma}^+$ (bottom) are produced.}
    \label{fig:dns3_snapshot9}
\end{figure}

Figure~\ref{fig:DNS3_flame_surface} displays the total flame surface in the domain versus time during DNS3.
Fig.~\ref{fig:dns3_snapshots} shows all the temporal snapshots of $c$ during DNS3, used for testing the CNN.
As the inlet speed is doubled, more mass flow enters the domain and the total flame surface increases. After the mass flow is set back to its initial value at snapshot $5$, the flame surface continues to increase until snapshot $\approx 9$, which matches the highly wrinkled aspect of the flame as seen in Fig.~\ref{fig:dns3_snapshot9}.
The mass flow then decreases below its original level, when the unburnt gas pocket exits the domain, starting at snapshot 15. The flame then grows back to its stable length and total area near snapshot 23. Snapshots after number 15 were not included in the testing dataset DNS3: indeed, no significant difference was observed, and this quasi-stable state is less challenging for the generalization of the trained network.

\begin{figure}[h!]
    \centering
    \includegraphics[width=0.5\linewidth]{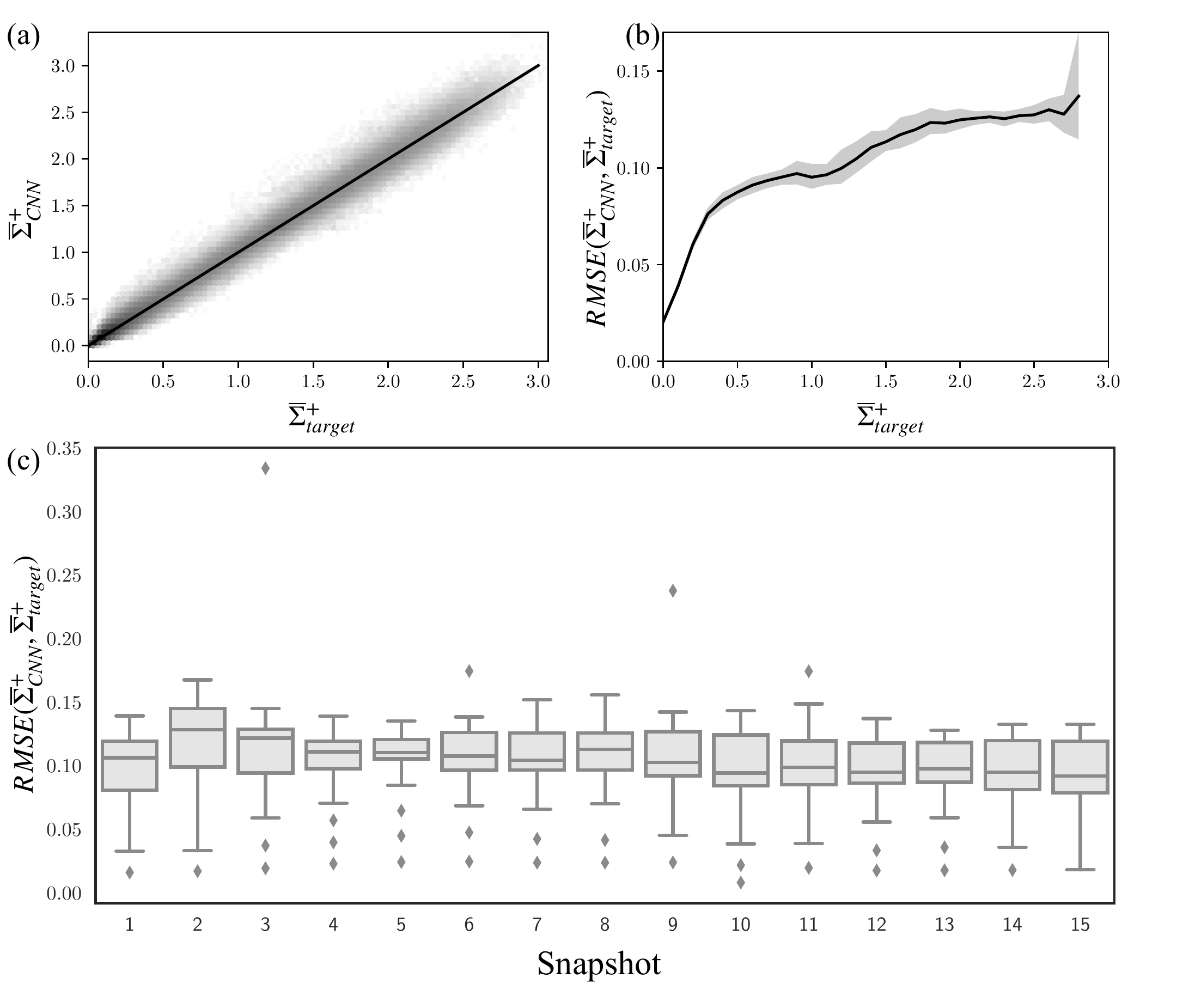}
    \caption{Data from the predictions of the CNN applied to the test dataset, \textit{i.e.} the 15 snapshots of DNS3. (a) Scatterplot of $\overline{\Sigma}^+$ predicted by the CNN vs target value extracted from the DNS. Grey line indicates $y = x$, the points with $0$ error. (b) Root Mean Squared Error (RMSE) of the $\overline{\Sigma}^+$ predictions versus target DNS value. Dark line is the mean value of RMSE over all $15$ snapshots, light-gray shows $95\%$ confidence interval around this value. (c) Boxplot view of RMSE distribution for each snapshot. Central bar: median value; grey box: first to last quartile; vertical line: $95\%$ occurrences; contour: kernel density plot estimate for full data.}
    \label{fig:DNS3_results}
\end{figure}

The objective of the network is to predict a value of $\overline{\Sigma}^+$ at every node and for every instantaneous snapshot that is as close as possible to the true value computed in the DNS $\overline{\Sigma}^+_{target}$. Fig.~\ref{fig:DNS3_results}~(a) shows the overall point by point agreement on the full test set, and demonstrates that the network recovers well the overall trend in the data. In order to better appreciate the error, Fig.~\ref{fig:DNS3_results}~(b) plots the Root Mean Squared Error (RMSE) of the prediction for bins of points sharing a predicted value in $0.1$-wide windows. This shows that the maximum RMSE occurs for the higher values of $\overline{\Sigma}^+$, and Fig.~\ref{fig:DNS3_results}~(c) indicates that some snapshots experience rare extreme RMSE values that can reach $0.4$. These events are however limited, and the majority of errors are in the $[0-0.2]$ range.  This is a normalized value directly comparable to $\overline{\Sigma}^+$, which is valued at $1$ in unwrinkled flame fronts and $\approx 2.5$ in highly wrinkled areas (Fig.~\ref{fig:dns3_snapshot9}).
From this we conclude that the transient data of DNS3 performs very well on the testing set in a statistical sense.

\subsection{Comparison with state of the art models}
In order to compare the method to the existing literature, a state of the art algebraic method was also implemented: the model of Charlette \textit{et al.}~\cite{Charlette:2002}, with a parameter value $\beta = 0.5$. This efficiency function assumes flame-turbulence equilibrium to evaluate the amount of sub-grid scale wrinkling, ultimately yielding $\Xi$. Equation~\ref{eq:wrinkling_factor} gives the relationship with $\overline{\Sigma}$, and therefore in order to compare this method to the output of the CNN the following field is built:
\begin{align}
    \overline{\Sigma}^+_{Charlette} = \frac{\Xi \lvert \nabla \overline{c} \lvert}{\overline{\Sigma}_{lam}^{max}}
    \label{eq:sigma_charlette}
\end{align}
with the same normalization factor as Eq.~\ref{eq:normalized_target}, $\overline{\Sigma}_{lam}^{max}$. Note that, as explained with Eq.~\ref{eq:charlette}, this form has been modified~\cite{Wang:2011} so that when ``saturated'' (\textit{i.e.} for high subgrid-scale velocities) it recovers the fractal model~\cite{Gouldin:1989} of Eq.~\ref{eq:gouldin}, with a fractal dimension of $D_f = 2.5$. The prediction made by this fractal model can also be compared to the present results in the form:
\begin{align}
    \overline{\Sigma}^+_{Fractal} = \frac{\Xi_{Fractal} \lvert \nabla \overline{c} \lvert}{\overline{\Sigma}_{lam}^{max}}
    \label{eq:sigma_gouldin}
\end{align}

\begin{figure*}[h!]
    \centering
    \includegraphics[width=\linewidth]{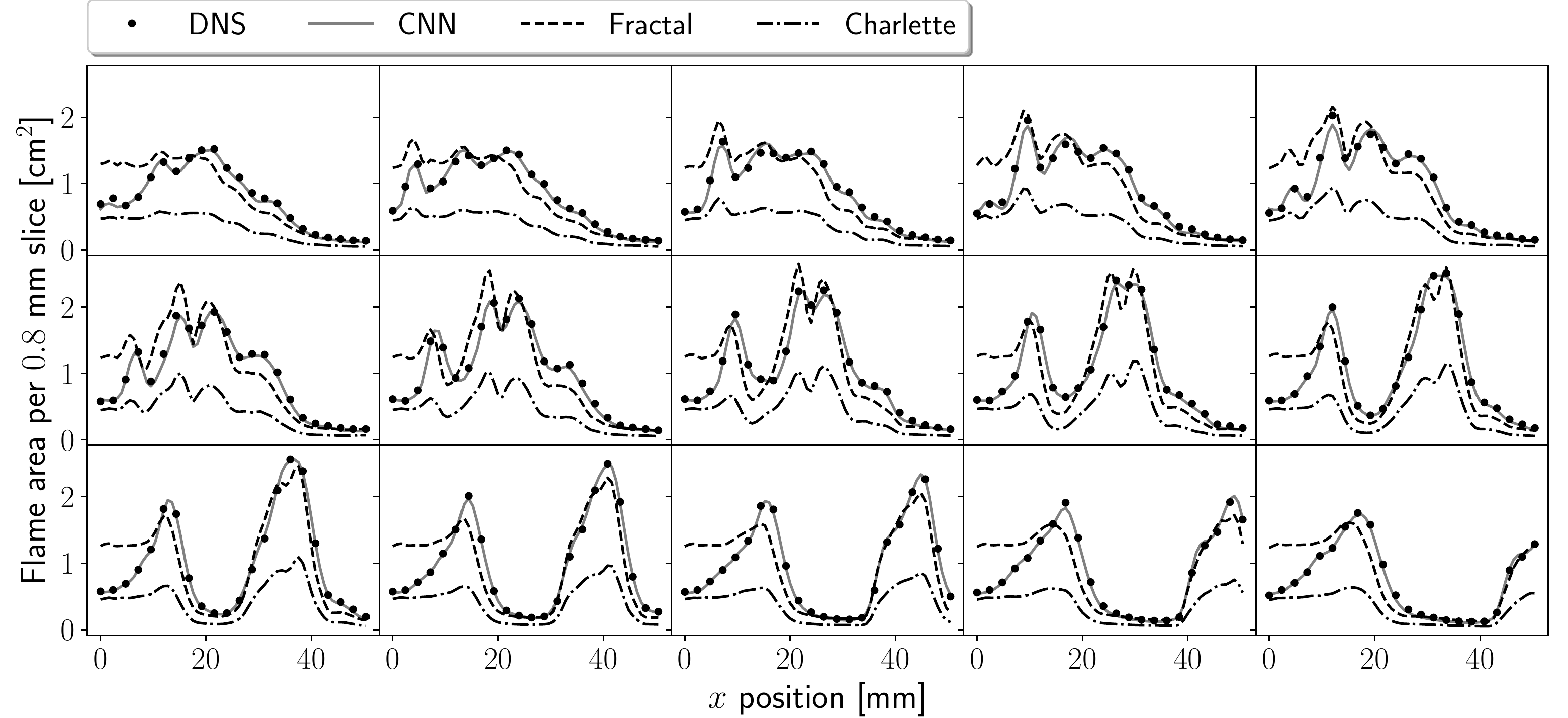}
    \caption{Total flame surface at each $x$ location for all snapshots of DNS3.}
    \label{fig:total_area_v_x_timeseries}
\end{figure*}

\begin{figure}[h!]
    \centering
    \includegraphics[width=0.5\linewidth]{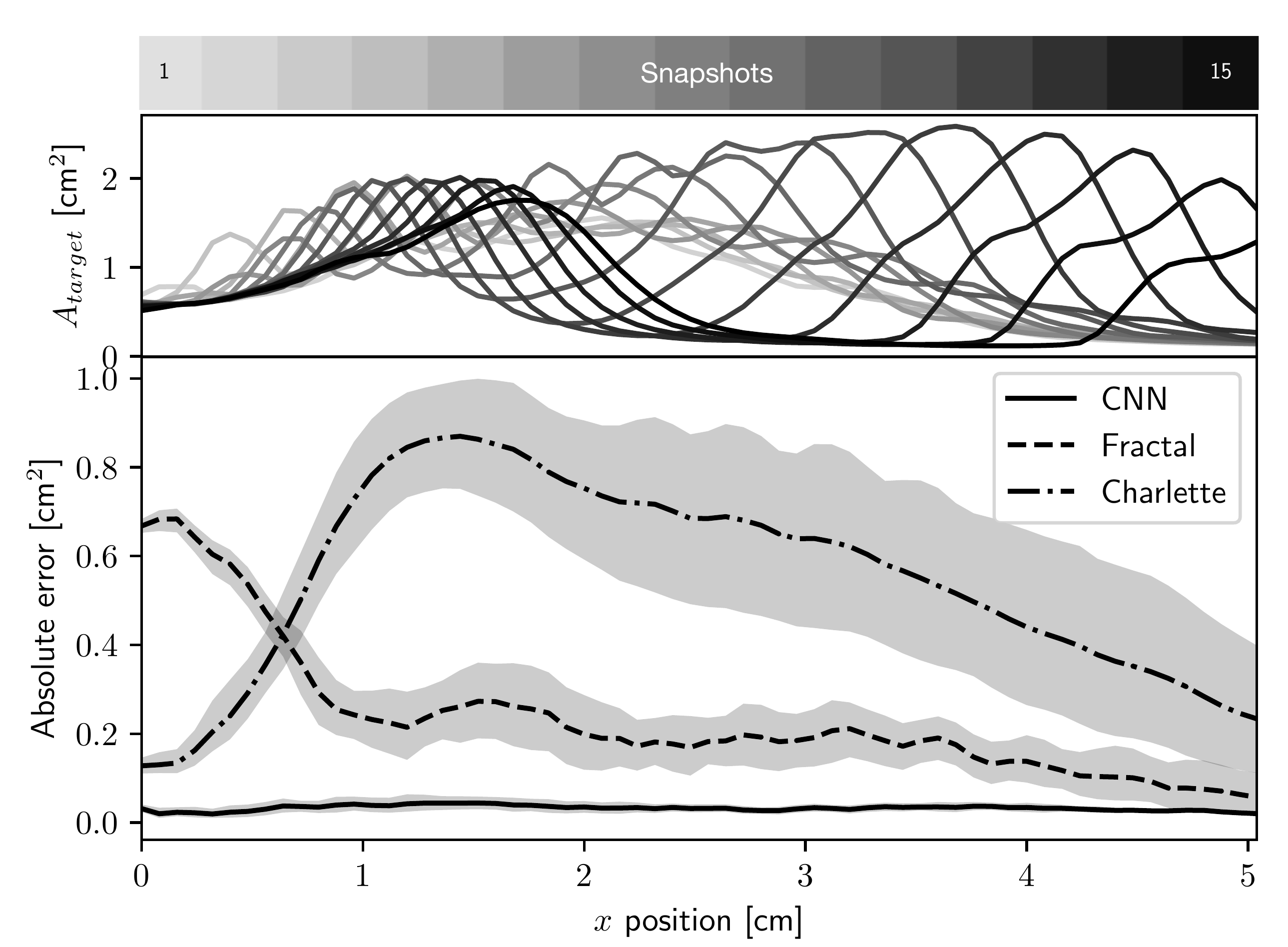}
    \caption{Top: Total flame surface at location $x$ (integrated in the $y-z$ plane) from DNS3 at each instant with snapshots ordered by increasing darkness; Bottom: Absolute error of CNN and Charlette predictions for the same $x$ locations. Grey area shows $95\%$ confidence interval.}
    \label{fig:error_vs_x}
\end{figure}

Fig.~\ref{fig:total_area_v_x_timeseries} displays the detailed views of the total flame area integrated in $8$ mm slices (1 LES mesh cell) along axial position for each snapshot of DNS3.
Fig.~\ref{fig:error_vs_x} displays the mean absolute prediction error of flame surface versus axial position $x$ for all DNS3 snapshots, \textit{i.e.} a condensed view of Fig.~\ref{fig:total_area_v_x_timeseries}.
It is clear from this graph that the CNN performs with very high accuracy compared to the other models. At the inlet, where turbulence is present but has not yet wrinkled the flame front, the fractal model overestimates significantly the amount of flame surface. This is due to the fact that the flame front is not yet wrinkled by turbulent structures. Turbulence and flame motions have not yet reached an equilibrium, usually assumed in the derivation of algebraic flame surface models. The CNN however has learned that the large-scale topology of the flow is not wrinkled, and therefore that the flame is in fact not wrinkled, leading to an excellent prediction in this region. Downstream, the Charlette model significantly underpredicts the flame surface (\textit{c.f.} Fig.~\ref{fig:total_area_v_x_timeseries}). This suggests that the Charlette model is far from its saturation value of Eq.~\ref{eq:gouldin}, hence the subgrid scale turbulence is underestimated in this configuration.

\section{Conclusions}
In this study, a convolutional neural network (CNN) inspired from a U-net architecture was used to predict sub-grid scale flame surface density for a premixed turbulent flame on an underresolved mesh typical of large eddy simulation.  Results show excellent agreement for this case between the prediction by the neural network and the baseline value from the DNS. Some algebraic models from the literature are shown for comparison and significantly under-perform compared to the CNN. This suggests, like the dynamic formulation has before, that including topological information to estimate unresolved flame wrinkling helps to achieve better accuracy. The advantage of the present technique is that this extraction was not hand-designed: with little effort, the training process leads to automatic and multi-scale extraction to assist the prediction.

The authors believe that this approach has demonstrated significant capacity to reach good accuracy in predicting subgrid scale contribution to flame wrinkling, as well as good capacity to generalize to a different test case than the training set. To the best of their knowledge, this is the first application of CNNs to turbulent combustion modeling.  This new technique poses some specific challenges, unlike traditional models, that will be investigated in future work to expand the capacity of the learned network to generalize to new configurations and flame regimes.  Indeed, while showing that the network can learn to reproduce the FSD on a given setup is a first step, the range of wrinkling values covered here is not absolute, and more variability should be introduced in the dataset in order to confirm that the network can in fact generalize to a broader array of cases.

\section*{Funding}

This research did not receive any specific grant from funding agencies in the public, commercial, or not-for-profit sectors.

\section*{References}

\end{document}